\documentclass[fleqn,10pt]{wlscirep}
\usepackage[utf8]{inputenc}
\usepackage[T1]{fontenc}
\title{Localization and Persistent Currents in a Quasiperiodic Disordered Helical Lattice}

\author[1]{Taylan Y\i ld\i z}
\author[1,*]{B. Tanatar}
\affil[1]{Department of Physics, Bilkent University, Ankara, 06800, T\"urkiye}

\affil[*]{email:tanatar@fen.bilkent.edu.tr}

\keywords{Helical lattice, Localization, Quasiperiodic disorder, Persistent Current}

\begin{abstract}
We investigate localization and persistent currents in a helical tight-binding lattice subject to two independent magnetic fluxes and a quasiperiodic on-site potential. Working with non-interacting, spinless fermions under periodic boundary conditions, we solve the model by exact diagonalization and study localization with both inverse and normalized participation ratios. We identify boundaries separating extended, mixed, and localized regimes by constructing a diagram incorporating potential strength and inter-ring coupling. In the metallic regime, persistent currents flowing around both the toroidal and poloidal directions show oscillations whose amplitude decays as disorder grows and vanishes past the localization threshold; in the localized regime, currents become flux-insensitive. We demonstrate that tuning magnetic fluxes, hopping strengths, or quasiperiodic potential amplitudes provides control over the critical disorder threshold. Our results suggest a versatile platform for disorder- and flux-controlled switching between conductive and insulating states. 
\end{abstract}
\begin{document}

\flushbottom
\maketitle

\thispagestyle{empty}

\section*{Introduction}
The study of localization and transport phenomena has attracted considerable attention for illuminating the interplay of disorder and geometry in quantum matter. Especially, for nanoscale electronics, photonic devices,
and quantum information platforms where controlled conduction is significant. 

The concept of localization in condensed matter physics started with Anderson localization (AL) \cite{1}, which showed that random uncorrelated potentials can localize electrons in a lattice and produce an insulating phase. 
This metal-insulator transition has also been studied experimentally in optical and photonic lattices \cite{2,3,4,5,6}. Similar phenomena can be obtained by replacing the random disorder in AL with a quasiperiodic potential in a one-dimensional tight-binding lattice. The so-called Aubry-Andr{\'e} (AA) model exhibits self-duality and a sharp metal-insulator transition at a critical modulation amplitude \cite{7,8}. Historically, the AA (Harper) model arises as the 1D Harper equation, which is obtained from the 2D Hofstadter problem \cite{33}: choosing a Landau gauge and Fourier transformation along a lattice direction reduces Bloch electrons in a uniform magnetic field to a 1D tight-binding chain with a cosine on-site potential. In recent years, a rich family of AA-type models has been explored, including those with next-nearest-neighbor or longer-range hopping, many-body extensions, dimerized lattices, non-Hermitian generalizations, and higher-dimensional analogues \cite{9,10,11,12,13,14,15,30,31,32}. 

Alongside the localization studies, the concept of persistent currents has provided a compelling window into quantum coherence and boundary condition effects. Persistent currents result from the Aharonov-Bohm (AB) effect because charged particles acquire a geometric phase proportional to the enclosed magnetic flux, even in regions where the magnetic field is zero \cite{16,17}. Theoretical studies have predicted that mesoscopic metallic rings can sustain nonzero currents whose magnitude and sign oscillate periodically with magnetic flux \cite{18}, a prediction that was later verified in experiments \cite{19,20}. Further studies have extended this idea to dimerized and topologically non-trivial lattices, where geometry, magnetic phases, and the localization phenomena intertwine in intricate ways \cite{21,22,23,24,25,26,27}.

This work presents a comprehensive study of a helical lattice constructed from rotated rings threaded by two independent magnetic fluxes \cite{21}, subject to a quasiperiodic AA modulation to enforce incommensurability across the torus. We begin by formulating a tight-binding Hamiltonian, describing the geometry and on-site modulation. Using exact diagonalization for non-interacting spinless fermions, we computed the inverse and normalized participation ratios (IPR and NPR) \cite{28} across a broad range of quasiperiodic potential strengths, inter-ring couplings, and 
magnetic fluxes. This allows us to construct the complete localization diagram, which includes extended, mixed, and localized regimes altogether. Additionally, we evaluated persistent currents circulating in both the toroidal and poloidal directions, uncovering flux-induced oscillations in the metallic regime and a suppression of oscillations beyond the localization threshold. Unlike a disorder-free helix and a randomly disordered (Anderson-type) helix, where the states are purely extended or sharply localized, the AA-type quasiperiodic potential offers a knob for controlling the amplitude of the currents while producing mobility edges (mixed regimes).

\section*{Model and Approach}
\begin{figure}[h]

\includegraphics[width=140mm]{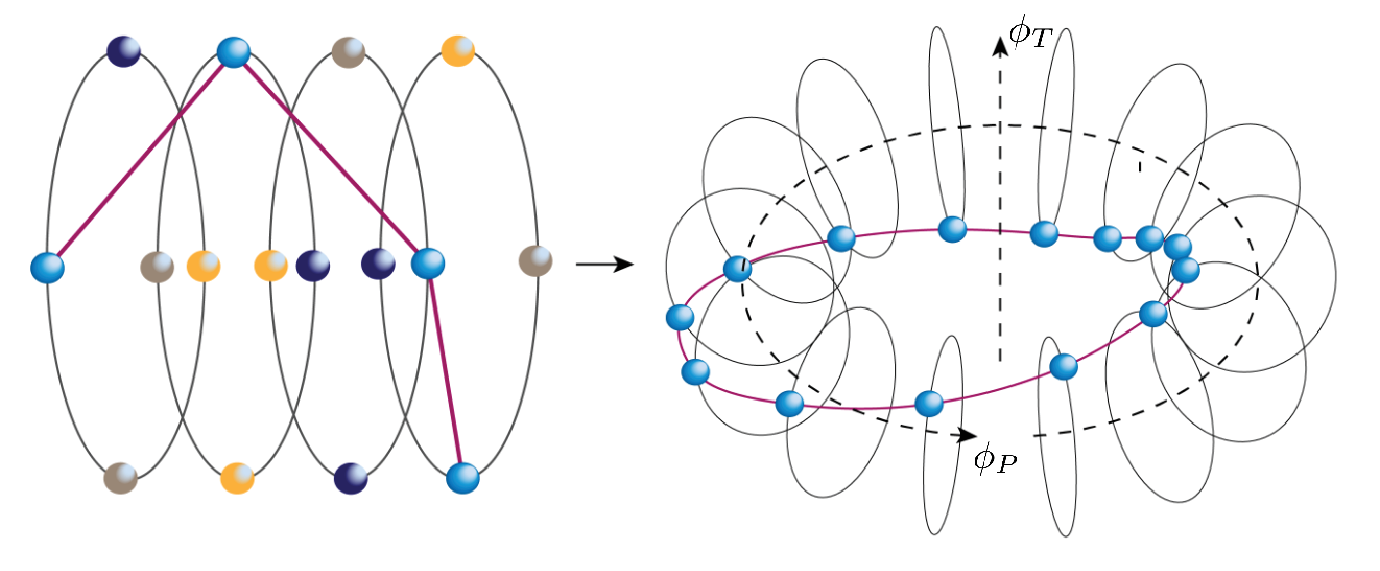}
\caption{(a) Schematic of the rotated-rings helical lattice. Each ring contains $N$ equally spaced sites and is rotated by $\Delta\theta=2\pi/L$ relative to its neighbor, forming a helical chain. Intra-ring hopping $J$ (solid black lines) connects nearest-neighbor sites around each ring, while inter-ring hopping $J_R$ (solid red line, only displayed through one site for clarity) links corresponding site positions between adjacent rings. (b) Imposing periodic boundary conditions along the ring loops and the stacking direction produces a toroidal geometry with two cycles. The magnetic flux $\phi_T$ encircles each ring loop (short cycle), while the flux $\phi_P$ threads vertically through the torus hole along the helical axis (long cycle).}\label{fig1}
\end{figure}
We consider a tight-binding model of non-interacting, spinless fermions on a helical lattice formed by $L$ rings. Each ring consists of $N$ sites uniformly spaced around its circumference.  As illustrated in Fig.\ref{fig1}, successive rings are rotated about the central axis at a fixed angle. This rotated-rings construction winds the rings into a single helical chain, producing the helical geometry of our model, initially introduced by Iskin et al. \cite{21}. 
\noindent The Hamiltonian reads as follows. 
\begin{align}
        H=-\sum_{p=1}^{L}\sum_{i=1}^NJe^{-i\alpha}c_{p,i}^\dagger c_{p,i+1}+h.c.-\sum_{p=1}^{L}\sum_{i=1}^NJ_Re^{-i\alpha N/L}&e^{-i\gamma}c_{p,i}^\dagger c_{p+1,i}+h.c.\nonumber\\&+\Delta\sum_{p=1}^{L}\sum_{i=1}^N\cos(2\pi\beta (i+Np)+\theta_p)c_{p,i}^\dagger c_{p,i}
\end{align}
Here, $c_{p,i}$ ($c^\dagger_{p,i}$) annihilates (creates) a spinless fermion on ring $p=1,\dots,L$ and site $i=1,\dots,N$ where $N$ is the number of sites per ring, $L$ the number of rings. The intra- and inter-ring hopping strengths are represented by $J$ and $J_R$, respectively. We impose periodic boundary conditions around each ring and along the stack of rings. Due to the magnetic fluxes $\phi_P$ and $\phi_T$ (see Fig.\ref{fig1}), phase factors appear
in the hopping parameters with $\alpha=2\pi\phi_P/N\phi_0$ and $\gamma=2\pi\phi_T/L\phi_0$ where $\phi_0= h/e$ is the flux-quantum. Since adjacent
rings are azimuthally offset by $\Delta\theta=2\pi/L$, stepping from ring $p$ to $p{+}1$ advances the azimuthal coordinate by $2\pi/L$. A complete $2 \pi$ turn around a ring introduces a phase $ e^{-iN \alpha}$; hence, an inter-ring hop picks up an additional phase $e^{-i(\alpha N/L)}$. The inter-ring hop therefore carries a phase $e^{-i(\gamma+\alpha N/L)}$, whereas the intra-ring hop carries $e^{-i\alpha}$. The two fluxes couple to different non-contractible cycles of the torus and are therefore topologically independent. They also appear linear in the phases on bonds, and thus can be realized independently, without interfering with each other. 
Finally, the on-site AA modulation is controlled by the strength $\Delta$, an incommensurate number 
$\beta=(\sqrt{5}-1)/2$ (inverse golden ratio) and a phase $\theta_p$. To ensure that every site of the torus experiences a unique quasiperiodic phase, we let the argument of the AA cosine term be $i+pN$. This way, on-site potential becomes incommensurate across the entire $L\times N$ lattice. 

Building such a toroidal helical lattice lets one control interference via two independent AB fluxes defined on the torus cycles, enabling flux-tunable localization thresholds and mesoscopic persistent currents in both directions while boosting the amplitude of the poloidal current with an additional phase on inter-ring bonds. A laboratory implementation of such connectivity can be realized by lithographed ring arrays on a cylindrical substrate using independent axial and toroidal (or saddle) coils to set up $\phi_T$ and $\phi_P$ or in cold-atom platforms where Raman-assisted tunneling separately controls the phases on the bonds \cite{34}. While a full helical torus with two AB fluxes hasn't been observed yet, we propose the key ingredients as: independent control of phases along the orthogonal directions in optical lattices \cite{35}, ring traps suitable for persistent currents \cite{36}, and an optical lattice with toroidal geometry \cite{37}. As an example to parameters, a lithographed ring of radius $r=0.5~\mu\mathrm{m}$ needs $B\simeq1.6~\mathrm{mT}$ to realize $\phi_T=0.3\,\phi_0$, while for the long-cycle (toroidal) flux threading the hole, a major radius $R=2~\mu\mathrm{m}$ requires only $B\simeq0.066~\mathrm{mT}$ to reach $\phi_P=0.2\,\phi_0$. In an optical lattice formed by a typical $1064~\mathrm{nm}$ laser (lattice spacing $a\approx 532~\mathrm{nm}$), our choices $\phi_P=0.2\,\phi_0$ and $\phi_T=0.3\,\phi_0$ correspond to Peierls angles $\alpha=2\pi \phi_P/N\phi_0\approx0.063$ and $\gamma=2\pi \phi_T/L\phi_0\approx0.094$ (for $N=L=20$), both can be achievable with Raman-assisted tunneling.

To quantify the localization properties of the system, we work with a single-particle solution of the model. Since we are 
using non-interacting spinless fermions, the full many-body eigenstates of the Hamiltonian can be decomposed into 
single-particle wave functions. Therefore, we can capture localization properties by examining the single-particle 
eigenstates of the Hamiltonian. Denoting the $n$th eigenstates by $\psi_n^{(i,p)}$ in the site basis, 
The IPR and NPR are defined as,
\begin{align}
    {\rm IPR}_n=\sum_{p=1}^L\sum_{i=1}^N|\psi_n^{(i,p)}|^4, \quad \quad {\rm NPR}_n=\left(LN\sum_{p=1}^L\sum_{i=1}^N|\psi_n^{(i,p)}|^4\right)^{-1}
\end{align}
To capture the overall localization of wave functions, we average IPR and NPR over all eigenstates, 
\begin{align}
    \langle {\rm IPR}\rangle=\frac{1}{LN}\sum_{n=1}^{LN}{\rm IPR}_n, \quad \quad \langle {\rm NPR}\rangle=\frac{1}{LN}\sum_{n=1}^{LN}{\rm NPR}_n
\end{align}
and also we use the composite quantity $\eta$ introduced in Ref. \cite{29}
\begin{align}
    \eta=\log_{10}(\langle {\rm IPR}\rangle\times\langle {\rm NPR}\rangle)
\end{align}
When the system is in the mixed regime both $\langle \rm IPR \rangle$ and $\langle \rm NPR \rangle$ are finite (i.e. $\sim\mathcal{O}(1))$ so the $\eta$ lies in $-2\lesssim\eta\lesssim-1$. However, if the system is in localized (extended) region $\langle \rm NPR \rangle$ ($\langle \rm IPR \rangle$) becomes $\sim LN^{-1}$ and we have $\eta<-\log_{10}NL$.

We promote our single-particle picture to a non-interacting many-fermion picture to analyze the persistent currents in a helical lattice. Then the ground-state energy is obtained by filling the lowest $N_f$ (number of fermions) single-particle energies,
\begin{align}
    E_g(\phi_P,\phi_T)=\sum_{n=1}^{N_f}\varepsilon_n(\phi_P,\phi_T)
\end{align}
Persistent currents in both toroidal and poloidal directions are defined as the first-order derivative with respect to the corresponding flux \cite{18}, 
\begin{align}
    I_P=-\frac{\partial E_g(\phi_P,\phi_T)}{\partial\phi_P}, \quad \quad I_T=-\frac{\partial E_g(\phi_P,\phi_T)}{\partial\phi_T}
\end{align}
These many-body currents quantify how the filled Fermi sea responds to variations in the two magnetic fluxes and provide a complementary transport diagnostic alongside single-particle localization measures.
\begin{figure}[h]

\includegraphics[width=140mm]{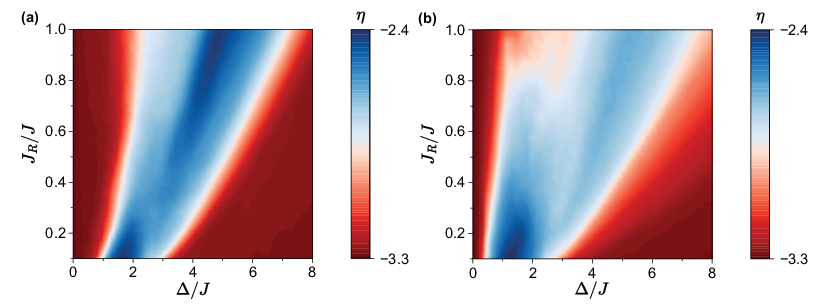}  
\caption{Color map of the composite localization metric $\eta$ plotted versus potential strength $\Delta/J$ and $J_R/J$ for a lattice size (a) $N=L=40$ (b) $N=30$, $L=60$. Red regions ($\eta\lesssim-3.2$) correspond to completely extended or localized states, whereas blueish regions ($-3.2\lesssim\eta$) correspond to mixed states. The mixed (blue) region expands with the increase of $J_R$ and produces energy-dependent localization thresholds (mobility edges).}\label{fig2}
\end{figure}

\section*{Results and Discussion}

This section displays the essential numerical results for localization and persistent current. We show the effects of hopping anisotropy, disorder strength, and applied flux on the interplay of conducting and insulating regimes. We scale energies by $J=1$ eV, and while reporting currents, we use the dimensionless ratio $I/I_0$ where $I_0=J/\phi_0=3.874\times10^{-5}\,$A. Unless stated otherwise, we average over the quasiperiodic phase $\theta_p$ and, when explicit values are required, take $\phi_T=0.3\phi_0$, $\phi_P=0.2\phi_0$, and $J_R=0.5J$.

\subsection*{Localization}

We present the result for localization in the form of a
color map in Fig.\ref{fig2}(a) for a lattice of size
$40\times40$. 
We used the composite metric $\eta$ for classification. The red regions correspond to homogeneous regimes in which the spectrum is either fully extended or localized (with negligible coexistence). The bluish region corresponds to mixed states, where localized and extended states coexist. As one can see, the two states have a sharp boundary. As $J_R$ increases, the mixed region shifts to higher values of potential strength $\Delta$. Larger inter-ring hopping $J_R$ broadens and hybridizes the minibands, which breaks the Aubry–André self-duality and generates mobility edges. Flatter subbands localize near the AA-like threshold, while more dispersive states remain extended up to larger $\Delta$. Therefore, the mixed region broadens with $J_R$ and starts to vanish when $\Delta$ is large enough to localize all subbands. In simpler words, stronger coupling between rings increases the connectivity of the lattice, making it harder for disorder to localize wave functions. The critical potential strength $\Delta_c$ to localize the helical lattice completely increases monotonically with increasing $J_R$. This trend shows us that the hopping strength $J_R$ gives us a clean handle to tune $\Delta_c$. Fig.\ref{fig2}(b) also indicates that the same localization scheme persists on a lattice of size $30\times60$, which shows us that boundary behavior on the $(J_R,\Delta)$ plane is robust to size changes. 
\begin{figure}[h]
	
	\includegraphics[width=90mm]{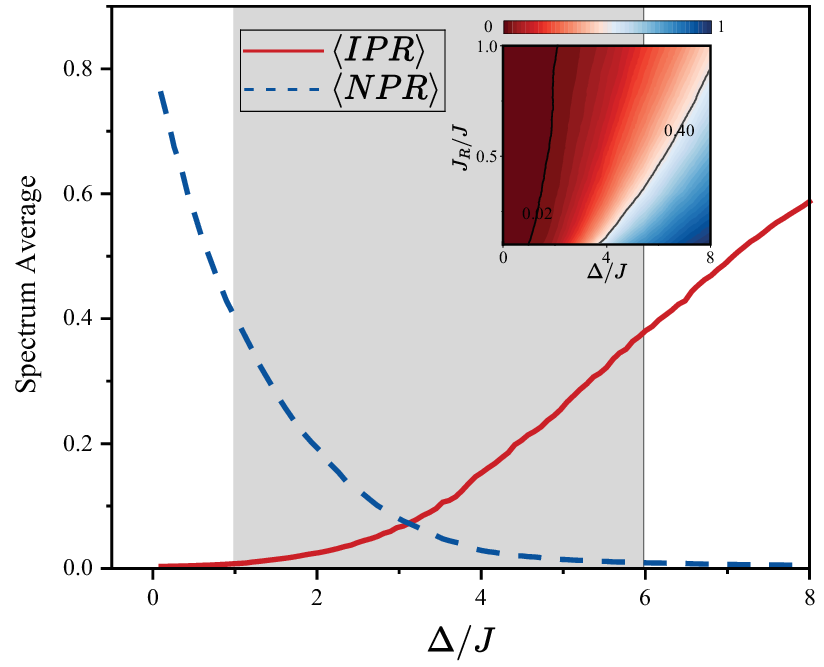}
	\caption{Spectrum-averaged IPR (solid red) and NPR (dashed blue) versus quasiperiodic potential strength $\Delta/J$ for $J_R=0.5J$ on a $20\times20$ lattice. The shaded region marks the mixed region. The inset shows the corresponding $\langle {\rm IPR}\rangle$ color map with respect to $J_R/J$ and $\Delta/J$; black contour lines correspond to the lower and upper edges of the mixed region. }\label{fig3}
\end{figure}
In Fig.\ref{fig3} we plot the spectrum-averaged $\langle {\rm IPR}\rangle$ (solid red) and $\langle {\rm NPR}\rangle$ (dashed blue) as functions of the quasiperiodic potential strength $\Delta$ for a fixed inter-ring coupling $J_R=0.5$ on a $20\times20$ lattice. At small $\Delta$, the vanishing $\langle {\rm IPR}\rangle$ and large $\langle {\rm NPR}\rangle$ indicate fully extended states; as $\Delta$ grows, both quantities become finite over a window (shaded gray) signaling the mixed region coexistence of localized and extended eigenstates. Beyond the upper edge of this gray band, $\langle {\rm IPR}\rangle$ saturates at high values while $\langle {\rm NPR}\rangle$ tends to zero, marking the onset of complete localization. The inset of Fig.\ref{fig3} repeats these boundaries in a two-dimensional color map of $\langle {\rm IPR}\rangle$ versus $(\Delta,J_R)$, with contour lines (e.g., 0.02 and 0.40) tracing the lower and upper localization thresholds on the same $20\times20$ lattice. Also, it is essential to note that since our study is mesoscopic, the boundaries in Figs. \ref{fig2},\ref{fig3} indicate crossovers between regions for the finite lattice, rather than sharp thermodynamic phase transitions.

We now focus on this $20\times20$ ($400$-site) system for our study of persistent currents because it is in the mesoscopic regime. The size is large enough to exhibit a clear localization transition in $\langle {\rm IPR}\rangle$, $\langle {\rm NPR}\rangle$, yet small enough to produce measurable nondecaying current oscillations. Larger lattices would suppress the amplitude of these oscillations below detectable levels, masking the mesoscopic transport that we wish to analyze \cite{17}.

Similarly, at the mesoscopic scale, the composite metric $\eta$ can become numerically unstable. Its logarithmic product of the IPR and NPR tends to wash out the details of localization when the total number of sites falls below $\sim10^3$. However, the individual averages $\langle {\rm IPR}\rangle$ and $\langle {\rm NPR}\rangle$ remain 
well-behaved. Hence, for our study of flux-driven persistent currents in this mesoscopic regime, we revert to separately tracking $\langle {\rm IPR}\rangle$ and $\langle {\rm NPR}\rangle$, which provide more reliable markers of the mixed and the extended or localized regions.

\begin{figure}[h]

\includegraphics[width=140mm]{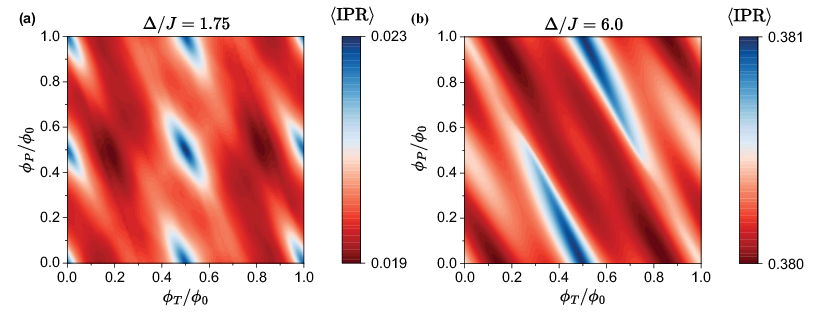}
\caption{$\langle {\rm IPR} \rangle$ on a $20\times20$ helical torus, shown as a function of the dimensionless flux $\phi_T/\phi_0$ (horizontal axis) and $\phi_P/\phi_0$ (vertical axis). (a) for $\Delta=1.75J$ (b) for $\Delta=6J$.}\label{fig4}
\end{figure}

Fig.\ref{fig4} illustrates how applied magnetic fluxes can impact localization. In the left panel (extended regime), we see $\langle {\rm IPR} \rangle$ exhibits clear ridges when either flux is at half a quantum ($\phi_T,\phi_P=0.5\;\phi_0$) or a full quantum ($\phi_T,\phi_P=\phi_0$). Each of the conditions maximizes destructive interference and localizes the states. By contrast, in the localized regime (right panel) $\langle {\rm IPR} \rangle$ is uniformly high across all flux values, showing that when the potential strength exceeds $\Delta_c$, magnetic flux has a negligible effect. This shows that tuning the flux can localize some states in the delocalized regime, but cannot drive the insulating regime back to a conducting one or vice versa. 

\begin{figure}[h]

\includegraphics[width=140mm]{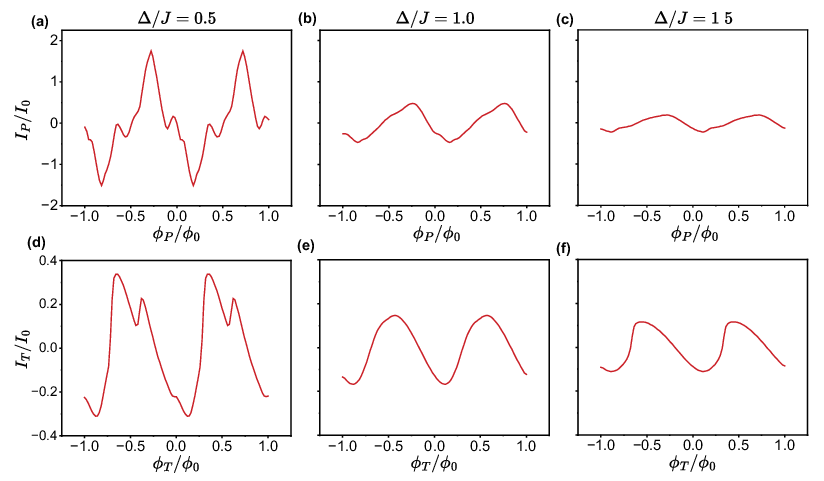}
\caption{Persistent currents on a $20\times20$ lattice ($J_R=0.5J$, half-filling) plotted against the corresponding dimensionless flux. Columns correspond to increasing on-site potential strength $\Delta$. $\Delta=0.5J$ for (a) and (d), $1.0J$ for (b) and (e), $1.5J$ for (c) and (f). The top row (a), (b), (c) shows the poloidal current $I_P/I_0$, and the bottom row (d), (e), (f) shows the toroidal current $I_T/I_0$.}\label{fig5}
\end{figure}

\subsection*{Persistent Current}
 This section presents our numerical results for equilibrium persistent currents in the rotated rings helix. Starting from the many-body ground state of $N_f$ non-interacting spinless fermions at half-filling, we compute the toroidal current $I_T$ and poloidal current $I_P$ as derivatives of the ground-state energy with respect to the corresponding dimensionless fluxes.

\begin{figure}[h]
\includegraphics[width=140mm]{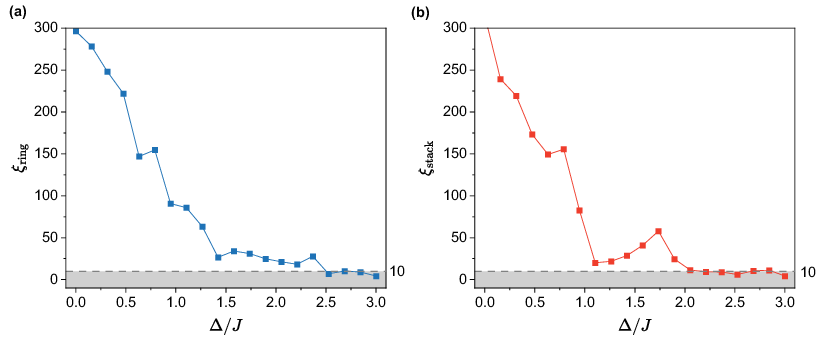}
\caption{Localization lengths (in lattice spacing units) (a) $\xi_{\rm ring}$ and (b) $\xi_{\rm stack}$ plotted against the quasiperiodic potential strength $\Delta/J$. Obtained by fitting the decay of the Fourier harmonics of the $E_g$. Parameters are $N=L=20$, $J_R=0.5J$, half-filling, and we swept across $512$ flux values each. The gray band marks the localized regime; $\xi_{\rm ring}\gg N/2$ and $\xi_{\rm stack}\gg L/2$ are effectively extended.  }\label{fig6}
\end{figure}

In Fig.\ref{fig5} we plot the variation of poloidal (upper row) and toroidal (bottom row) persistent currents with $\phi_P,\phi_T$ for evolving quasiperiodic potential strength $\Delta$ with columns. At $\Delta=0.5$ (first column), both $I_P$ and $I_T$ exhibit large, multi-harmonic oscillations, characteristic of the metallic regime with many conducting channels. At a small $\Delta$ (since we are in the extended regime), dozens of extended modes contribute their Fourier components to $E_g$, 
\begin{align}
    E_g(\phi_T,\phi_P)=\sum_{m=1}^{\infty}\sum_{n=1}^{\infty}A_{m,n}(\Delta)\cos(2\pi m\phi_T)\cos(2\pi n\phi_P)
\end{align}
and we have that \cite{18}
\begin{align}
    A_{m,n}\propto \exp{(-(m+n)C_{\text{loop}}/\xi(\Delta))}\equiv\exp(-mC_{\rm ring}/\xi_{\rm ring}(\Delta))\exp(-nC_{\rm stack}/\xi_{\rm stack}(\Delta))
\end{align}
Here $C_{\text{loop}}$ is the length of the closed loop around a torus cycle (in our lattice units; $C_{\rm ring}=N$ for the ring loop and $C_{\rm stack}=L$ for the stacking loop), $\xi(\Delta)$ is the single particle localization length \cite{7}. For a single–particle eigenstate $\psi$ on the torus,
the amplitude decays approximately as $|\psi(\mathbf r)|\!\sim\! e^{-d(\mathbf r)/\xi(\Delta)}$, with $\xi(\Delta)$
the localization length (On our torus, we resolve $d(\mathbf{r})$ along the two cycles as $d_{\rm ring}(i),\;d_{\rm stack}(p)$ and correspondingly use $\xi_{\rm ring}(\Delta),\;\xi_{\rm stack}(\Delta)$. As $\Delta$ increases, $\xi(\Delta)$ decreases rapidly, suppressing the higher harmonics $A_{m,n}$ and turning the multi-harmonic oscillations into nearly sinusoidal ones. To see the trend more clearly, we plot Fig.\ref{fig6}, which displays the localization lengths extracted and fitted from the decay of the Fourier harmonics of the ground state energy. We see the rapid drop of the lengths as the potential strength increases. The gray band identifies the localized region where $\xi_{\rm ring}< N/2$ and $\xi_{\rm stack}< L/2$ (since the geodesic distance on a periodic loop of length $N$ is at most $N/2$).

\begin{figure}[h]
\includegraphics[width=140mm]{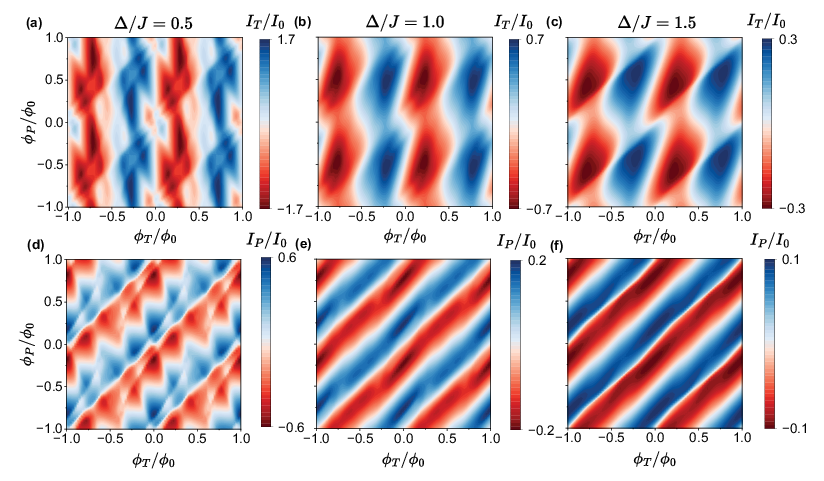}
\caption{Color maps of poloidal persistent current $I_P/I_0$ in (a), (b), (c) and toroidal persistent current $I_T/I_0$ in (d), (e), (f) as a function of $\phi_P/\phi_0$ (horizontal axis) and $\phi_T/\phi_0$ (vertical axis). The three panels correspond to increasing potential strengths $\Delta$. $\Delta=0.5J$ for (a) and (d), $1.0J$ for (b) and (e), $1.5J$ for (c) and (f).}\label{fig7}
\end{figure}
The local extrema in $I_{P,T}$ arise from the constructive and destructive interference of Fourier components. Gradual disappearance with $\Delta$ tracks the loss of extended states as identified by the drop of $\xi_{\rm ring}$ and $\xi_{\rm stack}$ in Fig.\ref{fig6}. Hence, in the completely extended  (Fig.\ref{fig5}, first column), the resulting current is a superposition of many harmonics and exhibits sharp oscillations. In Fig.\ref{fig5}, at $\Delta=1.0$ (second column), the amplitude of the currents is roughly halved, and the waveforms have smoothed out since there are fewer contributing Fourier coefficients, reflecting a partial localization. Finally, at $\Delta=1.5$ (third column), the current amplitude becomes weaker and nearly harmonic, signaling more eigenstate localization.

The poloidal current $I_P$ drops faster than the toroidal current $I_T$, consistent with the earlier collapse of the $\xi_{\rm stack}$ than the $\xi_{\rm ring}$. The difference comes from the form of the on-site potential. When stepping around the successive sites on a ring, the argument of the cosine shifts by $2\pi\beta$; however, stepping around the successive rings shifts the cosine by $2\pi\beta N$, which is lower than $2\pi\beta$ in our $20\times20$ lattice. The fast modulation around ring sites leads to stronger back-scattering on the bonds.

In Fig.\ref{fig7} for continuity, we present the color maps of poloidal current $I_P$ and toroidal current $I_T$ over the full two-flux plane for three representative potential strengths. We see similar effects as we discussed with the Fig.\ref{fig5}. These maps directly visualize how increasing the quasiperiodic potential strength systematically filters out higher-order flux oscillations and suppresses the magnitude of the currents.

\begin{figure}[b]
\includegraphics[width=140mm]{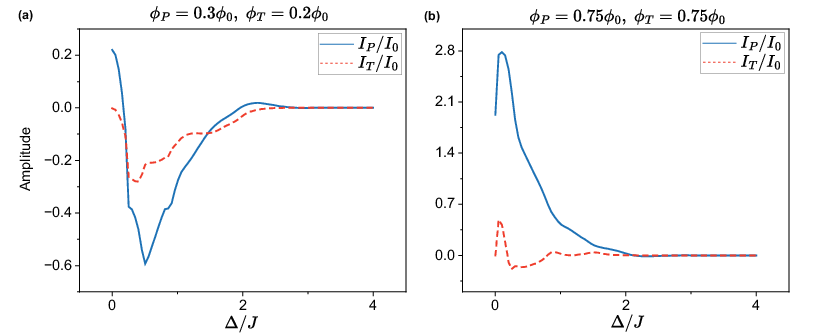}
\caption{Poloidal $I_P/I_0$ and toroidal $I_T/I_0$ persistent currents plotted versus quasiperiodic potential strength $\Delta/J$. Magnetic fluxes are (a) $\phi_P=0.3\phi_0,\;\phi_T=0.2\phi_o$ (b) $\phi_P=0.75\phi_0,\;\phi_T=0.75\phi_o$.}\label{fig8}
\end{figure}

Additionally, Fig.\ref{fig8} shows how the amplitudes of $I_P$ and $I_T$ evolve as the quasiperiodic potential $\Delta$ increases for two different flux configurations. Both of the currents display non-monotonic dependence on $\Delta$. In a disorder-free helical lattice ($\Delta=0$), the single particle bands are perfectly sinusoidal and symmetric in momentum \cite{21}. With half-filling, the bands just above and below the Fermi level cancel each other due to perfect lattice and particle-hole-like symmetries (opposite k-slopes). However, turning on a small potential modulation perturbs the canceling states and enhances the persistent current, producing the maximum current amplitude. Further increasing $\Delta$, the localization length $\xi$ shrinks, high-order harmonics are filtered, the system goes from a completely extended regime to a partially extended regime, and the current amplitudes are suppressed to nearly sinusoidal traces of small magnitude. Beyond a critical quasiperiodic strength $\Delta_c$, the system enters a localized (insulating) regime, and the persistent currents $I_T$ and $I_P$ vanish for our finite lattices. This quench also correlates with the mixed-to-localized crossover seen in $\langle\mathrm{IPR}\rangle$\;\text{and}\;$\langle\mathrm{NPR}\rangle$. 
We can also observe from Fig.\ref{fig8} that the localization of the many-body ground state $E_g$ happens at 
$\Delta\approx2.5$  because the single-particle states below the Fermi level have all crossed into the localized region. We see that Fig. \ref{fig3} shows that the upper edge of the mixed window at $J_R=0.5$ lies near $\Delta\approx6$. This mismatch comes from the fact that the Fig. \ref{fig3} tracks localization across the entire spectrum. However, in Fig. \ref{fig8}, we only track the localization of the lowest $N_f$ states. The persistent currents depend only on the occupied single–particle states, since the localization threshold is energy dependent in mixed region (mobility edges), the $\Delta$ at which $E_g$ becomes flux–insensitive is the largest critical value across the occupied window, $\Delta_{\rm occ}(f)=\max_{E\le E_F(f)}\Delta_c(E)$ where $E_f$ is the Fermi energy. So the point that current vanishes
shifts with filling: if $E_F$ lies near band edges, occupied states localize at smaller $\Delta$; if $E_F$ sits in highly dispersive subbands, larger $\Delta$ is required. Our results at half filling and $J_R=0.5$ show that the occupied window localizes around $\Delta\approx2.5$,
even though Fig. \ref{fig3} shows that unoccupied high–energy bands remain extended up to $\Delta\approx6$. Similarly, the localization length analysis in Fig.\ref{fig6} supports this: as $\Delta$ increases, the extracted lengths $\xi_{\rm ring}$ and $\xi_{\rm stack}$ collapse and cross the threshold at $\Delta \simeq 2.5$ (slightly earlier in the stack direction), so that the filled single–particle states are localized along both torus cycles. Beyond this point, the current harmonics are suppressed, so $E_g$ becomes flux–insensitive and the persistent currents vanish.

\section*{Conclusion}
Our research has demonstrated that a helical lattice with a quasiperiodic potential enables precise control over the transition between insulating and conducting regions, while simultaneously enabling control over persistent currents in two distinct directions. The same parameters that govern localization also determine the behavior of these currents. Weak quasiperiodic modulation enhances current flow, while stronger modulation suppresses it through localization of eigenstates. By tuning the external magnetic flux and finely adjusting transport channels, one can shift the strength of the critical disorder. These findings not only deepen our theoretical understanding but also open new avenues for designing flux- and disorder-controlled insulating devices or waveguides. Looking ahead, it will be important to include interactions. Adding on-site Hubbard $U$ interactions should renormalize the localization length and potentially narrow the mixed window or probe many-body localization at large $\Delta$. The geometry of the model also suggests practical implementations for future experimental realizations in terms of nanofabricated ring arrays wrapped on a cylinder with axial and circumferential coils, helical photonic waveguide lattices with synthetic Peierls phases, and toroidal cold-atom lattices using Raman-assisted tunneling. These directions,  combined with finite-size scaling, are expected to complement the structure to push flux- and disorder-controlled devices towards practical applications.

\section*{Data Availability}
Data sets generated during the current study are available from the corresponding author on reasonable request.

\bibliography{references}

\section*{Figure Legends}
\begin{enumerate}
    \item (a) Schematic of the rotated-rings helical lattice. Each ring contains $N$ equally spaced sites and is rotated by $\Delta\theta=2\pi/L$ relative to its neighbor, forming a helical chain. Intra-ring hopping $J$ (solid black lines) connects nearest-neighbor sites around each ring, while inter-ring hopping $J_R$ (solid red line, only displayed through one site for clarity) links corresponding site positions between adjacent rings. (b) Imposing periodic boundary conditions along the ring loops and the stacking direction produces a toroidal geometry with two cycles. The magnetic flux $\phi_T$ encircles each ring loop (short cycle), while the flux $\phi_P$ threads vertically through the torus hole along the helical axis (long cycle).
    \item Color map of the composite localization metric $\eta$ plotted versus potential strength $\Delta/J$ and $J_R/J$ for a lattice size (a) $N=L=40$ (b) $N=30$, $L=60$. Red regions ($\eta\lesssim-3.2$) correspond to completely extended or localized states, whereas blueish regions ($-3.2\lesssim\eta$) correspond to mixed states. The mixed (blue) region expands with the increase of $J_R$ and produces energy-dependent localization thresholds (mobility edges).
    \item Spectrum-averaged IPR (solid red) and NPR (dashed blue) versus quasiperiodic potential strength $\Delta/J$ for $J_R=0.5J$ on a $20\times20$ lattice. The shaded region marks the mixed region. The inset shows the corresponding $\langle {\rm IPR}\rangle$ color map with respect to $J_R/J$ and $\Delta/J$; black contour lines correspond to the lower and upper edges of the mixed region.
    \item $\langle {\rm IPR} \rangle$ on a $20\times20$ helical torus, shown as a function of the dimensionless flux $\phi_T/\phi_0$ (horizontal axis) and $\phi_P/\phi_0$ (vertical axis). (a) for $\Delta=1.75J$ (b) for $\Delta=6J$.
    \item Persistent currents on a $20\times20$ lattice ($J_R=0.5J$, half-filling) plotted against the corresponding dimensionless flux. Columns correspond to increasing on-site potential strength $\Delta$. $\Delta=0.5J$ for (a) and (d), $1.0J$ for (b) and (e), $1.5J$ for (c) and (f). The top row (a), (b), (c) shows the poloidal current $I_P/I_0$, and the bottom row (d), (e), (f) shows the toroidal current $I_T/I_0$.
    \item Localization lengths (in lattice spacing units) (a) $\xi_{\rm ring}$ and (b) $\xi_{\rm stack}$ plotted against the quasiperiodic potential strength $\Delta/J$. Obtained by fitting the decay of the Fourier harmonics of the $E_g$. Parameters are $N=L=20$, $J_R=0.5J$, half-filling, and we swept across $512$ flux values each. The gray band marks the localized regime; $\xi_{\rm ring}\gg N/2$ and $\xi_{\rm stack}\gg L/2$ are effectively extended.
    \item Color maps of poloidal persistent current $I_P/I_0$ in (a), (b), (c), and toroidal persistent current $I_T/I_)$ in (d), (e), (f) as a function of $\phi_P/\phi_0$ (horizontal axis) and $\phi_T/\phi_0$ (vertical axis). The three panels correspond to increasing potential strengths $\Delta$. $\Delta=0.5J$ for (a) and (d), $1.0J$ for (b) and (e), $1.5J$ for (c) and (f).
    \item Poloidal $I_P/I_0$ and toroidal $I_T/I_0$ persistent currents plotted versus quasiperiodic potential strength $\Delta/J$. Magnetic fluxes are (a) $\phi_P=0.3\phi_0,\;\phi_T=0.2\phi_o$ (b) $\phi_P=0.75\phi_0,\;\phi_T=0.75\phi_o$.
\end{enumerate}

\section*{Acknowledgments}
We thank B. Hetenyi for his helpful comments.

\section*{Funding}
This work was supported in part by the Turkish Academy of Sciences (TUBA).

\section*{Author contributions statement}

B.T conceptualized the project, T.Y. conducted numerical calculations and analyzed the data, T.Y. and B.T. co-wrote the paper. 

\section*{Additional information}
\textbf{Competing interests}  
The authors have no competing interests to declare relevant to the content of this article.

\end{document}